\documentclass[12pt,preprint]{aastex61}

\usepackage{natbib,amssymb,amsmath,graphicx}
\usepackage{textcomp}
\usepackage{longtable}

\begin{document}
\title{V1224 Cas - an EL CVn-type Eclipsing Binary Consisting of a Helium White Dwarf Precursor and a Delta Scuti Pulsator}

\correspondingauthor{Kun Wang}
\email{kwang@cwnu.edu.cn}

\author{Kun Wang}
\affiliation{Department of Astronomy, China West Normal University, Nanchong 637002, China}

\author{Changqing Luo}
\affiliation{Key Laboratory of Optical Astronomy, National Astronomical Observatories, Chinese Academy of Sciences, Beijing 100012, China}

\author{Xiaobin Zhang}
\affiliation{Key Laboratory of Optical Astronomy, National Astronomical Observatories, Chinese Academy of Sciences, Beijing 100012, China}

\author{Bo Zhang}
\affiliation{Key Laboratory of Optical Astronomy, National Astronomical Observatories, Chinese Academy of Sciences, Beijing 100012, China}

\author{Licai Deng}
\affiliation{Department of Astronomy, China West Normal University, Nanchong 637002, China}
\affiliation{Key Laboratory of Optical Astronomy, National Astronomical Observatories, Chinese Academy of Sciences, Beijing 100012, China}

\author{Zhiquan Luo}
\affiliation{Department of Astronomy, China West Normal University, Nanchong 637002, China}

\begin{abstract}
We report the discovery of a new eclipsing EL CVn-type binary, consisting of a $\delta$ Sct-type pulsator and a thermally bloated low-mass pre-He white dwarf (WD). Spectroscopy and time-series $BV$ photometry of V1224 Cas were carried out. The spectroscopy reveals a spectral type of A3 for the star. Light-curve modelling indicate that V1224 Cas is a short-period detached system containing a possible low-mass WD with an effective temperature of about 9516 K. Based on the effective temperature and the surface gravity of the A-type primary star from the spectroscopic results, the absolute parameters of the components were estimated as: $M_{P}$=2.16$\pm$0.22$M_{\sun}$, $R_{P}$=3.54$\pm$0.12$R_{\sun}$, $L_{P}$=55.9$\pm$6.9$L_{\sun}$, $M_{S}$=0.19$\pm$0.02$M_{\sun}$, $R_{S}$=0.97$\pm$0.04$R_{\sun}$, and $L_{S}$=6.9$\pm$0.6$L_{\sun}$. We therefore introduce V1224 Cas as a new EL CVn-type binary candidate. The light curves in both filters all show multi-periodic pulsations, superimposed on binary effects. We performed a preliminary frequency analysis of the light residuals after removing the synthetic eclipsing curve from the original observational data. The results suggest that the rapid light variations among the light curves could be attributed to the $\delta$ Sct-type primary component. We therefore conclude that V1224 Cas is very likely a WD+$\delta$ Sct binary. 
\end{abstract}


\section{Introduction}
Low-mass white dwarf  stars (WDs; $M$$\lesssim$$0.45M_{\sun}$), which likely harbor a He core, are generally thought to be the product of strong mass-loss episodes during the red giant branch phase of low-mass stars in interactive binary systems before the occurrence of He-flash \citep{mar1995, alt2013, ist2016, che2017}. When the mass transfer ends, the donor star still has a thick hydrogen envelope surrounding the helium core. The helium white dwarf precursor (pre-He-WD hereinafter) then evolves to higher effective temperature at nearly constant luminosity through an active hydrogen burning shell. Recently, dozens of pre-He-WDs were found in EL CVn-type binaries \citep{max2014a,van2018}, which consist of an A/F type main sequence star and a pre-He-WD ($M$$\sim$0.15-0.33$M_{\sun}$). \citet{max2014a} discovered 17 EL CVn systems using the SWASP photometric database \citep{pol2006} and chose the brightest one, EL CVn,  as the prototype of this class of eclipsing binaries. A total of 13 such samples were discovered in the Kepler survey. They are KOI 74, KOI 81 \citep{row2010,van2010}, KIC 10657664 \citep{car2011}, KOI 1224 \citep{bre2012}, KIC 9164561, KIC 10727668 \citep{rap2015}, KIC 4169521,  KOI-3818, KIC 2851474, KIC 9285587\citep{fai2015}, KIC 8262223 \citep{guo2017}, KIC 10989032 and KIC 8087799 \citep{zha2017}. By applying machine learning techniques, \citet{van2018} discovered 36 eclipsing EL CVn binaries \textbf{using data from the Palomar Transient Factory}. 

Among the observed EL CVn-type binaries, several of them have been shown to pulsate. Multi-periodic pulsations had been detected on three very-low-mass pre-He-WDs in the EL CVn-type systems WASP 0247-25 \citep{max2013}, WASP 1628+10 \citep{max2014b} and KIC 9164561\citep{zha2016}. Not only the pre-He-WD companion, but the dwarfs can also show pulsations. Five pulsating EL CVn-type binaries were reported \citep{max2014b, fai2015, guo2017, zha2017}, each consisting of a pre-He-WD and a $\delta$ Scuti-type pulsator. The $\delta$ Scuti stars are pulsating main sequence or less-evolved stars with masses in the range 1.5-2.5$M_{\sun}$, situated in the extensive part of the Cepheid instability strip towards low luminosities \citep{bre2000}.  Both radial and nonradial oscillations, driven by the $\kappa$ mechanism, occur in $\delta$ Scuti stars with typical periods in the range 12 min to 8 hr \citep{aer2010}. To date there are about 200 $\delta$ Scuti stars discovered in eclipsing binaries \citep{lia2017}. Those with short orbital periods, such as oscillating eclipsing systems of Algol type (oEA stars, \citealt{mkr2004}),  very probably undergo mass-transfer episodes resembling the EL CVn-type systems. Some of them may eventually evolve into exotic system containing a $\delta$ Sct-type pulsator plus a low-mass pre-He-WD companion. Such systems that have both binary properties and asteroseismology are interesting objects for exploring the stellar interiors and evolution.

In this work, we introduce V1224 Cas as a new EL CVn-type binary candidate containing a $\delta$ Sct-type pulsator. V1224 Cas (=UCAC4 733-108496, $\alpha_{J2000} = 23^{h}57^{m}38.59^{s}$, $\delta_{J2000} = +56^{\circ}35'57''.80$) is located in the field of the open cluster NGC 7789. The light variability of V1224 Cas was first noticed by \citet{noc1999}. Based on their sparse data, the authors classified this star as an EA-type eclipsing binary with an orbital period of 2.1077 days. But up to now, pulsating characteristics of V1224 Cas has not been reported.

As a contribution to the ongoing project on search for and study of variable stars in the 50BiN open cluster survey \citep{wan2015b}, we have performed time-series CCD photometry of the open cluster NGC 7789. V1224 Cas was detected in the program field. With intensive observational data and careful analysis, we found that the star could very likely be a sixth EL CVn-type binary consisting of a low-mass pre-He-WD and a $\delta$ Scuti-type pulsator. We report the discovery in this paper.

\section{Observations and data reduction}
The photometric observations were made with the 50 centimeter binocular telescope at the Qinghai Station of Purple Mountain Observatory \citep{den2013,tia2016}. This telescope, as the prototype of the 50 cm Binocular Network (50BiN), has two parallel camera systems. Each of them has a 2k $\times$ 2k Andor CCD camera and a set of standard Johnson/Cousins $UBVRI$ filters. The pixel scale is about 0.59 arcsec/pixel and the field of view is $\sim $20 $\times$ 20 arcmin$^2$. Two filters $B$ and $V$, fixed on the two tubes respectively, were applied to simultaneous two-color photometry. V1224 Cas has been monitored on 21 nights (a total of $\sim $135.2 hours) from 2015 August 24 to October 14. An overview of the time-series photometric observations is given in Table 1.  

\begin{deluxetable}{lcccc}
\tablecolumns{5}
\tablewidth{0pc}
\tablecaption{Log of time-series photometric observations for NGC 7789}
\tablehead{\colhead{Date} &\colhead{Start Date}     &\colhead{Length} & Exposure time   &\colhead{Number of Frames}\\
\colhead{} &\colhead{(HJD 2457250+)}     &\colhead{(hr)}             &\colhead{($B, V$) (in seconds) }    &\colhead{($B, V$)}}
\startdata
2015 Aug 24        &  9.116          &6.5       &140, 70      &137, 315\\
2015 Aug 29        &14.116          &2.7       &100, 50      &52, 154\\
2015 Aug 31        &16.065          &6.6       &140, 70      &125, 282\\
2015 Sep 04        &20.265          &3.1       &140, 70      &57, 150\\
2015 Sep 11        &27.128          &4.1       &140, 70      &87, 139\\
2015 Sep 13        &29.049          &8.5       &140, 70      &184, 411\\
2015 Sep 14        &30.048          &8.4       &140, 70      &180, 401\\
2015 Sep 16        &32.135          &6.3       &140, 70      &128, 300\\
2015 Sep 18        &34.152          &6.0       &140, 70      &123, 260\\
2015 Sep 30        &46.043          &7.3       &140, 70      &93, 360\\
2015 Oct  01        &47.027          &8.0       &140, 70      &164, 359\\
2015 Oct  03        &49.018          &7.7       &140, 70      &172, 369\\
2015 Oct  04        &50.015          &4.5       &140, 70      &88  , 198\\
2015 Oct  06        &52.014          &8.2       &140, 70      &187, 385\\
2015 Oct  07        &53.152          &4.8       &140, 70      &115, 229\\
2015 Oct  08        &54.026          &7.7       &140, 70      &178 , 370\\
2015 Oct  09        &55.015          &7.6       &140, 70      &168 , 351\\
2015 Oct  10        &56.112          &5.4       &140, 70      &122,  260\\
2015 Oct  12        &58.076          &6.2       &140, 70      &147,  300\\
2015 Oct  13        &59.005          &7.8       &140, 70      &181,  376\\ 
2015 Oct  14        &60.001          &7.8       &140, 70      &180,  380\\ 
\enddata
\end{deluxetable}

The photometric reductions were performed by using an automated reduction pipeline \citep{wan2015b}. It primarily includes: bias subtraction, flat-field correction, astrometric calibration and photometry extraction. Since we did not make observations of standard stars, the instrumental magnitudes of stars were standardized using 45 secondary standard stars in NGC 7789. Their standard magnitudes $B$ and $V$ were taken from Stetson Standard Star Catalog\footnote{http://www.cadc-ccda.hia-iha.nrc-cnrc.gc.ca/en/community/STETSON/standards/} \citep{ste2000}. A detailed procedure was described in our earlier work \citep{wan2015b}.

The instrumental magnitudes of time-series $BV$ frames were calibrated with an ensemble normalization technique \citep{gil1988}. Following the same procedure used in \citet{wan2015b}, about 45 isolated standard stars were picked out for the calibration. Using the following normalization equations, the instrumental magnitudes ($b$, $v$) were transformed to the magnitudes ($B$, $V$) on the standard Johnson system. 

\begin{equation}
B = b + a_{1} + a_{2}(B - V) +a_{3}X +a_{4}Y,
\end{equation}
 
\begin{equation}
V = v + c_{1} + c_{2}(B - V) +c_{3}X +c_{4}Y,
\end{equation}

Where $X$ and $Y$ are the star positions on a CCD frame. The coefficients for each CCD frame $a_{1}$/$c_{1}$, $a_{2}$/$c_{2}$, $a_{3}$/$c_{3}$ and $a_{4}$/$c_{4}$ were calculated by the least-squares method with the 45 standard stars. 

In order to make a better classification for the star, additional spectroscopic observations at low resolution were taken on 2017 September 15 with the BFOSC instrument attached to the 2.16m telescope at Xinglong Station of the National Astronomical Observatories, Chinese Academy of Sciences \citep{fan2016}. A low-resolution grism with a slit width of 1.$\arcsec$8 (G6 + filter of 385LP) was used during the observations. The spectra covers the wavelength range from 3300 to 5450 $\AA$ with a nominal resolving power of $\sim$2020. The unprocessed frames were reduced following the standard CCD procedure in the IRAF package \citep{tod1986,tod1993}. The CCD reductions include mainly bias subtraction and flat field correction. And then the wavelength calibration was carried out using the Fe/Ar lamp. In Figure 1, we plotted the normalized spectrum of V1224 Cas and the best 3 matches with the MILES standard spectra \citep{san2006}. It presents a typical feature of A-type stars. The best match of the spectrum of the system was found with that of an A3V standard star. The average effective temperature $T_{eff}$ and  surface gravity $log$ $g$ of the best 3 matches are 8395.0 $\pm422$ K and 3.9 $\pm0.24$ dex, respectively. 

\begin{figure}
\center
\includegraphics[scale=.9]{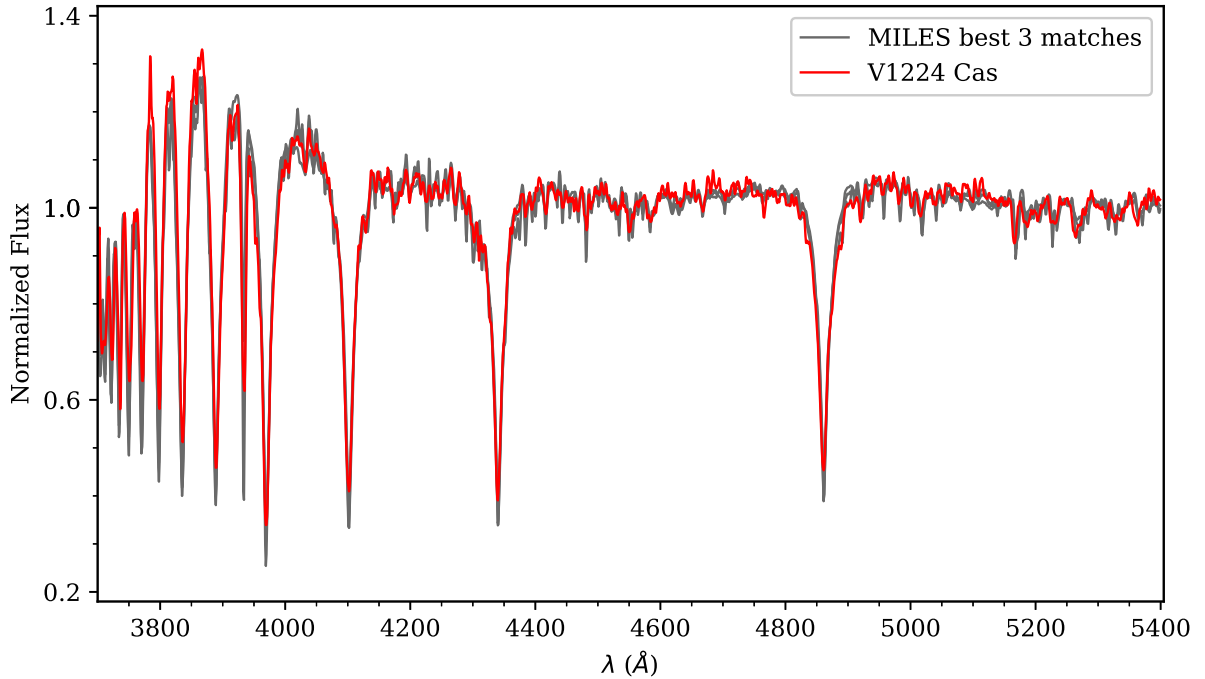}
\caption{Normalized spectrum of V1224 Cas (red line) matching with 3 MILES standard spectra (dimgray line), whose average effective temperature and surface gravity are 8395.0 $\pm422$ K and 3.9 $\pm0.24$ dex, respectively.}
\end{figure}

\section{Light curve modeling and system parameters}
 The original real-time $B-$ and $V-$band light curves of V1224 Cas are displayed in Figure 2. A total of 7 eclipse-like events were recored in this observational season. Not only the eclipse-like light variations, but the light curves can be also seen to show short-term pulsations, with cycle length and amplitude characteristic of $\delta$ Scuti-like oscillations. Since there are not enough minimum light times available, we employed the phased dispersion minimization (PDM) method \citep{ste1978} to look for the orbital period of the binary system. The orbital period of V1224 Cas was calculated to be 2.27537$\pm0.00001$ days, apparently longer than the result (2.1077 days) given by \citet{noc1999}. The phases of all the measurements were computed with the reference epoch ($T_{0}$=2457308.17 days) and the newly derived orbital period. The phase-folded light curves are shown in the upper panel of Figure 3. The light curves of V1224 Cas all present a flat-bottomed primary eclipse and a slightly shallower secondary eclipse, similar to the EL CVn-type binaries \citep{max2014a,van2018}. The depths of the two eclipses are estimated to be about 0.135 and 0.102 mag, respectively. These characteristics indicate that V1224 Cas may be a new EL CVn-type system with multi-periodic pulsations.
 
\begin{figure}
\center
\includegraphics[scale=0.9]{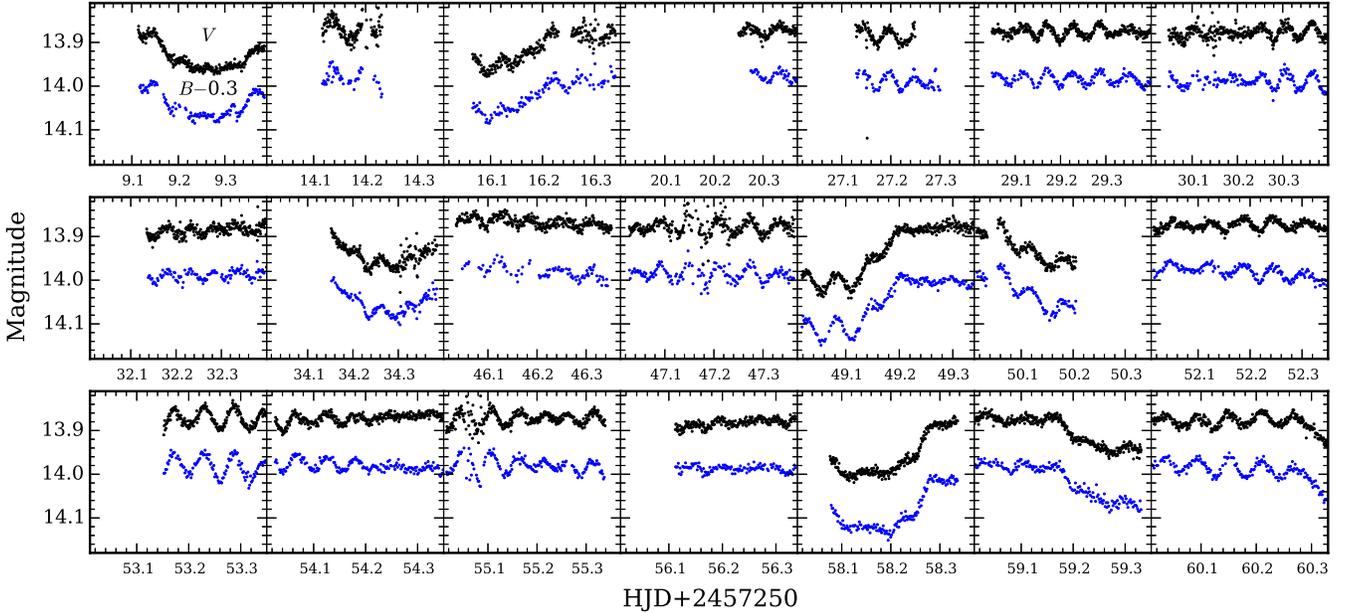}
\caption{The original real-time $B-$ and $V-$band light curves of V1224 Cas. The $B-$band points have been shifted by -0.3 mag in order to show light-curve features more clearly.}
\end{figure}

\begin{figure}
\center
\includegraphics[scale=1.]{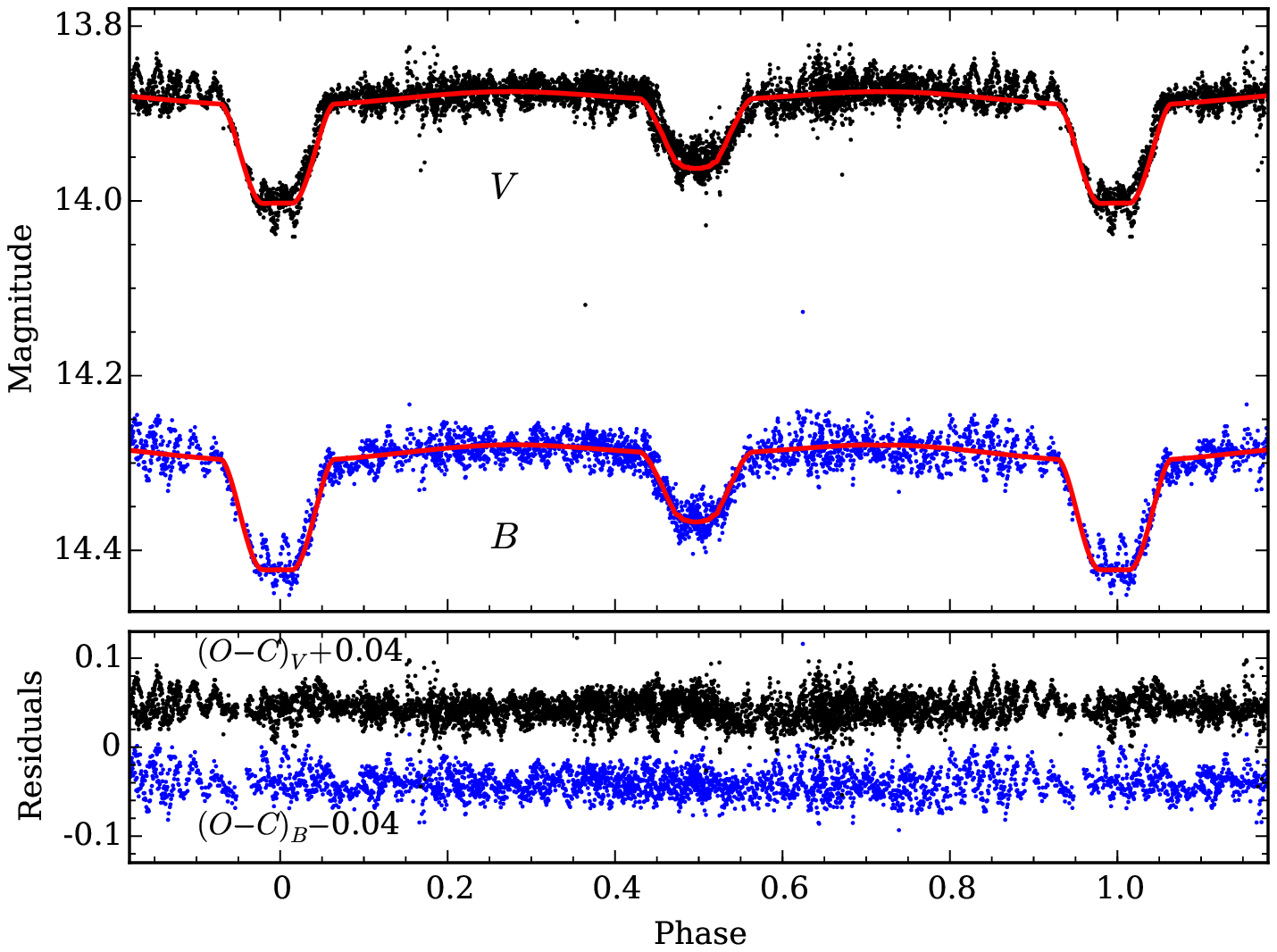}
\caption{Top panel: phase-folded light curves of V1224 Cas compared to the theoretical synthesis (red solid lines). Bottom panel: the O-C residuals of light curves calculated as observed minus the theoretical value.}
\end{figure}

To test the above hypothesis and further clarify the pulsating properties of this binary system, our two-color light curves were simultaneously analyzed by applying the 2013 version of the Wilson-Devinney (W-D) binary code \citep{wil1971,wil1979,wil1990,wil2012b}. The light-curve synthesis was carried out in a similar way to that for the eclipsing binaries KIC 9164561 \citep{zha2016}, HH UMa \citep{wan2015a} and V410 Aur \citep{luo2017}. We designated the luminous primary component as star 2 and its surface temperature is fixed at $T_{2}$=8395 K from the former spectroscopic result. The corresponding bolometric limb-darkening coefficients ($X_{1}$, $X_{2}$, $Y_{1}$, $Y_{2}$) and monochromatic ones ($x_{1}$, $x_{2}$, $y_{1}$, $y_{2}$) were interpolated using the values from \citet{van1993}'s tables with a logarithmic law. As V1224 Cas is \textbf{an} A-type star whose envelopes are probably in radiative equilibrium, we therefore adopted the theoretical values of the gravity-darkening exponents ($g_{1}$=$g_{2}$=1.0, \citealt{luc1967}) and the bolometric albedos ($A_{1}=A_{2}=1.0$, \citealt{ruc1969}). The adjustable parameters in the binary model are the mean surface temperature of star 1 ($T_{1}$), the phase shift, the orbital inclination ($i$), the mass ratio ($q=M_{2}/M_{1}$), the dimensionless potential of both components ($\Omega$$_{1}$, $\Omega$$_{2}$), and the monochromatic luminosity of star 1 ($L_{1B}$, $L_{1V}$). The subscripts 1 and 2 in this paper represent the two components being eclipsed at orbital phases 0.0 and 0.5, respectively.

Owing to no spectroscopic radial velocity measurements available for V1224 Cas, we first applied a photometric $q$-search method \citep{zha2015} to look for an approximate mass ratio. Then it was taken as a free parameter to be adjusted along with other adjustable parameters. Table 2 shows the final results from the best-fitting solution. Based on that information, we made the theoretical light curve and geometric configuration of V1224 Cas using the W-D binary model. We plotted the synthetic light curves (red solid lines) and the O-C residuals (observations minus calculations) in Figure 3. 

The photometric solution suggests a detached configuration for V1224 Cas. The geometric configurations of the binary system at phases 0.0, 0.25, 0.50, and 0.75 are illustrated in Figure 4. It is impossible to calculate directly the physical parameters of the binary system due to no spectroscopic orbital element. Following \citet{lee2017}, the mass of the primary component was estimated to be 2.16$\pm$0.22 $M_{\sun}$ using an empirical relation of the mass as a function of the observed $T_{eff}$, $log$ $g$ and [Fe/H] \citep{tor2010}, with an error of 10$\%$ assumed. The mass of 0.19$\pm$0.02 $M_{\sun}$ for the secondary component can be easily obtained from the photometric mass ratio ($q=M_{2}/M_{1}$). The semimajor axis of the binary system was calculated to be 9.67$\pm$0.33 $R_{\sun}$ according to the Kepler's third law. The luminosity and radius of the two components were then computed and given in the bottom of Table 2. The bolometric corrections (BCs) corresponding to the effective temperature of each component were derived from the relationship between $log$ $T$ and BC \citep{flo1996}. With an interstellar absorption of $A_{V}$=0.868 mag. \citep{wu2007} and an apparent magnitude of $V_{max}$=13.857 mag., the distance modulus $(m-M)_{o}$ of the system was estimated to be about 12.6 mag. \textbf{This star appears in the Gaia DR2 catalogue with a parallax of 0.2869$\pm$0.0236 mas \citep{bro2018}. Based on the parallax, the distance modulus of V1224 Cas is calculated to be about 12.71 mag, which agrees very well with our previous estimate. It shows that the previous assumptions made in the analysis are reasonable. As a result, V1224 Cas is  not a member of NGC 7789, since it is too large compared with the distance of the open cluster (see Table 1 in \citealt{wu2007}).}

\begin{deluxetable}{lccc}
\tablecolumns{4}
\tablewidth{0pc}
\tablecaption{Photometric solutions and physical parameters of the binary system V1224 Cas}
\tablehead{\colhead{Parameter} &   \colhead{Component 1}           &System        &\colhead{Component 2} }
\startdata
$P_{orb}$ (days)                          &\nodata                                      &2.27537$\pm0.00001$ &\nodata\\
$i$ (degree)                                 &\nodata                                       &76.3$\pm$0.1             &\nodata\\
$q=M_{2}/M_{1}$                         &\nodata                                       &11.47$\pm$0.07         &\nodata\\    
$T_{eff}(K)$                                 &9516$\pm$26                             &\nodata                       &8395$^{a}$\\
$\Omega$                                    &21.549$\pm$0.038                     &\nodata                       &27.829$\pm$0.045\\
$L_{i}/(L_{1}+L_{2})_{B}$             &0.072$\pm$0.003                       &\nodata                       &0.928$\pm$0.003\\
$L_{i}/(L_{1}+L_{2})_{V}$             &0.068$\pm$0.002                       &\nodata                       &0.932$\pm$0.003\\
$r$ (pole)                                      &0.0989$\pm$0.0005                   &\nodata                       &0.3581$\pm$0.0016\\
$r$ (point)                                     &0.1015$\pm$0.0005                   &\nodata                       &0.3713$\pm$0.0019\\
$r$ (side)                                      &0.0996$\pm$0.0005                    &\nodata                      &0.3678$\pm$0.0018\\
$r$ (back)                                     &0.1012$\pm$0.0005                    &\nodata                      &0.3698$\pm$0.0019\\
Absolute parameters:                   &                                                   &                                  &\\
Semimajor axis ($R_{\sun}$) &\nodata                                        &9.67$\pm$0.33          &\nodata   \\
$M$ ($M_{\sun}$)                         &0.19$\pm$0.02                            &\nodata                      &2.16$\pm$0.22 \\ 
$R$ ($R_{\sun}$)                          &0.97$\pm$0.04                            &\nodata                      &3.54$\pm$0.12\\
$L$ ($L_{\sun}$)                           &6.9$\pm$0.6                                &\nodata                      &55.9$\pm$6.9\\                                         
\enddata
\tablecomments{$^a$ Fixed; $L_{i}/(L_{1}+L_{2})$ and $r$ refer to the fractional luminosity and the equivalent radii of the two component stars, respectively.}
\end{deluxetable}

\begin{figure}
\center
\includegraphics[scale=0.9]{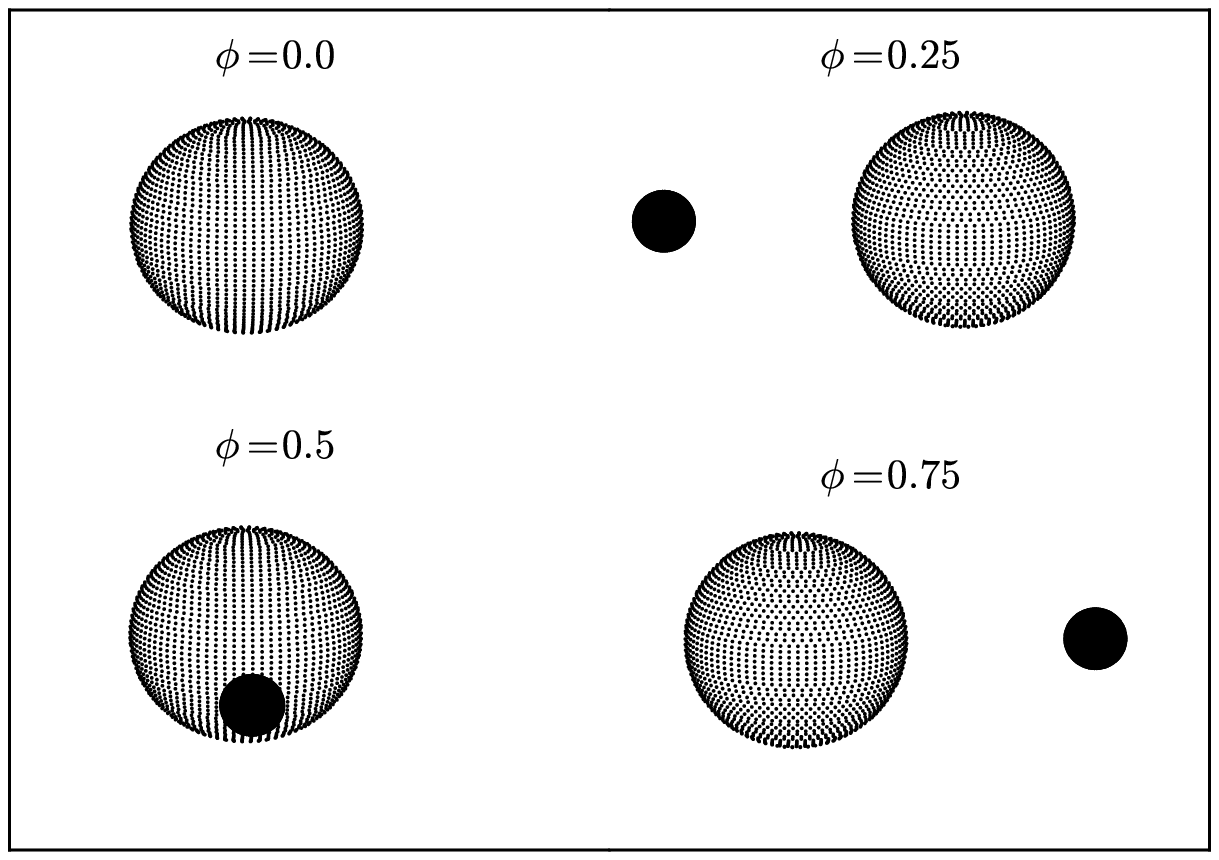}
\caption{The geometric configurations of V1224 Cas at phases 0.0, 0.25, 0.5, and 0.75.}
\end{figure}

\section{Light residuals and frequency analysis}
The light-curve residuals, obtained after subtracting the eclipsing light variations from the real-time light curves, are displayed in Figure 5 in the form of the magnitude versus HJD time instead of orbital phase. Wherein, multi-periodic light variability of this star are clear. The peak-to-peak magnitude of the light variations is estimated to be about 60 mmag. To investigate the pulsation nature of V1224 Cas, we performed a frequency analysis on the $B$- and $V$-band residual light curves using the software package Period04 v.1.2 \citep{len2005}. The pre-whitening technique was also employed for consecutive detection of periodic signals in the light residuals. Only those peaks that have a signal-to-noise amplitude ratio (S/N) larger than 4.0 in both filters were picked out for further analysis (following \citealt{bre1993}). Table 3 gives the main results of frequency analysis. The amplitudes and phases of all detected frequencies were determined by a non-linear, least-squares fitting routine in Period04. Their uncertainties were calculated with the relations reported by \citet{mon1999}. We computed the noise levels on the basis of the residuals from the original data after pre-whitening all the trial frequencies. The fitting curves to the original $B$- and $V$-band light curves were then made and displayed in Figure 5 with red solid lines. It indicates that the synthetic light curves adequately describe the observed light variations. In figure 6, we plot the spectral windows and step-by-step amplitude spectra of both the $B$- and $V$-band data, wherein each spectrum panel was calculated based on the residuals that all the previous frequencies were pre-whitened. 

The general features of the periodograms are typical of $\delta$ Scuti stars with multi-periodicity. Five significant peaks were detected in both the $B$- and $V$-filter data, with frequencies from 15.45 to 25.80 c/d and semi-amplitudes between 7.5 and 1.4 mmag. We examined the frequencies for possible combination ($f_{i}$ + $f_{j}$ or $f_{i}$ + $f_{orb}$) or harmonic terms ($Nf$). The differences of $f_{3}$ - $f_{1}$= 2.1749 c/d, $f_{4}$ - $f_{1}$= 2.682 c/d, and  $f_{5}$ - $f_{2}$= 6.5595 c/d are close to the values of 5$f_{orb}$=2.1975 c/d, 6$f_{orb}$=2.6369 c/d, and 15$f_{orb}$=6.5924 c/d, respectively. All of this suggests then that the three frequencies $f_{3}$, $f_{4}$, and $f_{5}$ are unlikely to be eigenmodes, but splitting components of $f_{1}$ and $f_{2}$ related to the orbital frequency $f_{orb}$=0.43949 c/d, respectively. The three frequencies $f_{3}$, $f_{4}$, and $f_{5}$ are excluded from the following discussion. 

\begin{figure}
\center
\includegraphics[scale=0.9]{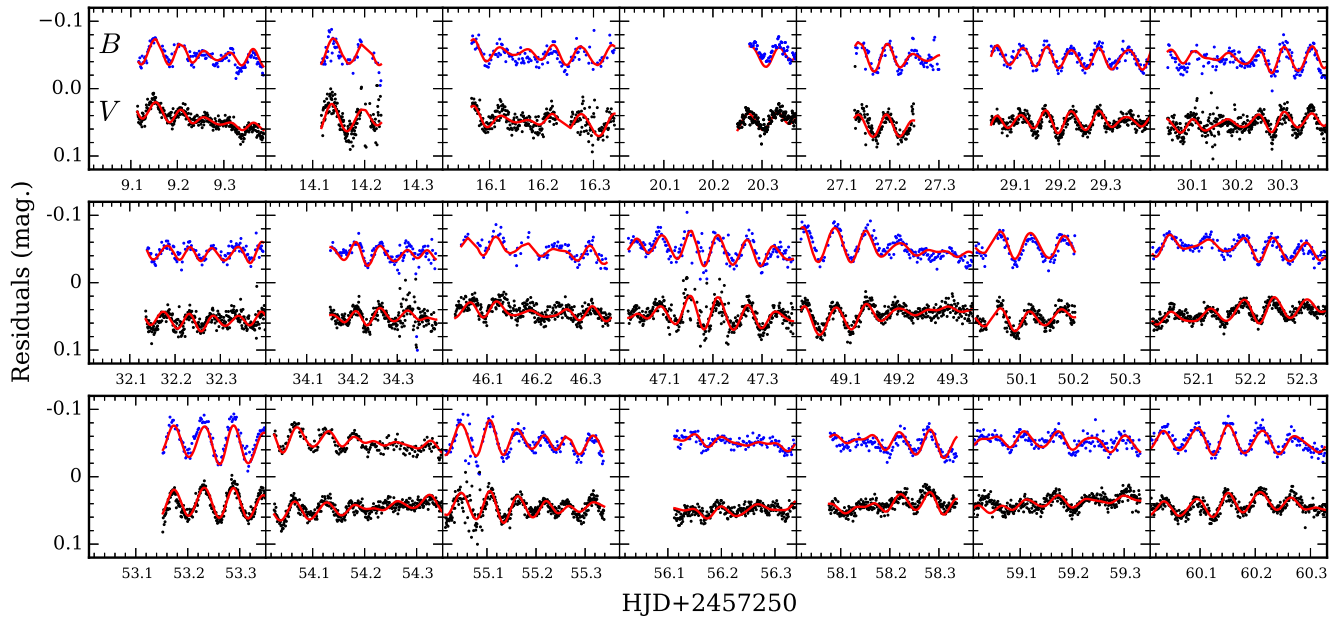}
\caption{Light curve residuals of V1224 Cas after removing the binary curve and Fourier fitting using the five-frequency in Table 3.}
\end{figure}

\begin{deluxetable}{lccccc}
\tablecolumns{6}
\tablewidth{0pc}
\tablecaption{Results of the Fourier analysis of the residual light curves.}
\tablehead{\colhead{Band}          &ID             &\colhead{Frequency (c/d)}     &\colhead{Amplitude (mmag.)}       &Phase             &\colhead{S/N}}
\startdata
$B$                                          &$f_{1}$     &15.4516$\pm0.0004$            &7.5$\pm$0.3                                  &0.385$\pm$0.006  &14.5\\
                                                 &$f_{2}$     &19.2354$\pm0.0005$            &6.8$\pm$0.3                                  &0.860$\pm$0.007  &9.6\\
                                                 &$f_{3}$     &17.6267$\pm0.0005$            &6.3$\pm$0.3                                  &0.833$\pm$0.007  &10.3\\
                                                 &$f_{4}$     &18.1342$\pm0.0009$            &3.5$\pm$0.3                                  &0.964$\pm$0.014  &6.7\\
                                                  &$f_{5}$     &25.7949$\pm0.0013$            &2.5$\pm$0.3                                  &0.516$\pm$0.019  &5.6\\
$V$                                           &$f_{1}$     &15.4516$\pm0.0004$            &5.9$\pm$0.2                                  &0.390$\pm$0.006  &9.8\\
                                                  &$f_{2}$     &19.2347$\pm0.0004$            &5.5$\pm$0.2                                  &0.081$\pm$0.006  &9.8\\
                                                  &$f_{3}$     &17.6263$\pm0.0004$            &5.4$\pm$0.2                                  &0.954$\pm$0.006  &10.4\\
                                                  &$f_{4}$     &18.1330$\pm0.0008$            &2.9$\pm$0.2                                  &0.306$\pm$0.012  &6.2\\
                                                 &$f_{5}$     &25.7943$\pm0.0016$            &1.4$\pm$0.2                                  &0.668$\pm$0.024  &4.8\\
\enddata
\end{deluxetable}

\begin{figure}
\center
\includegraphics[scale=1.]{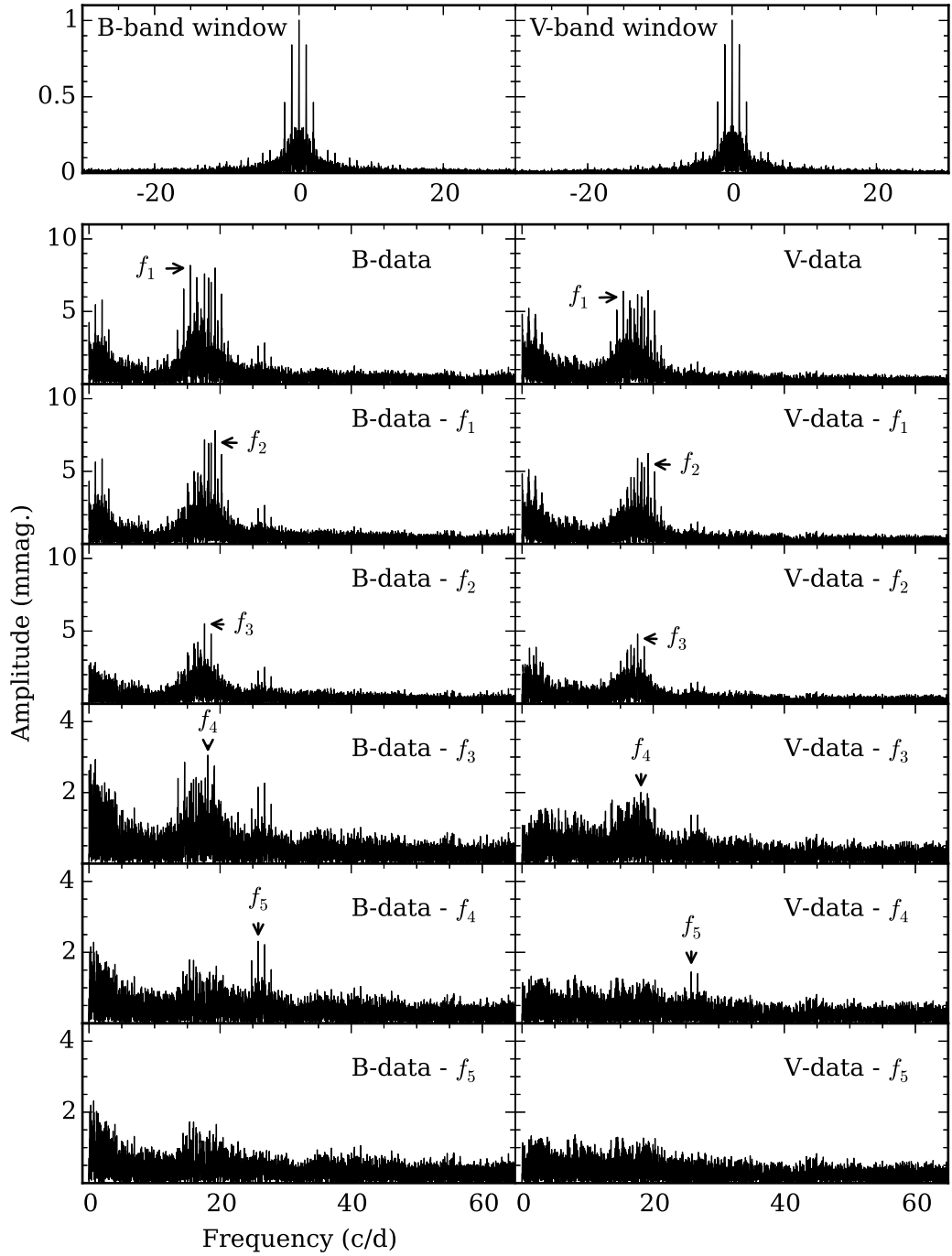}
\caption{Spectral windows and the step-by-step amplitude spectrum of V1224 Cas after subtracting the final binary model from the original light curves.}
\end{figure}

\section{Summary and Conclusion}
We have presented the time-series photometry and low-resolution spectroscopy of the eclipsing binary V1224 Cas. Based on which, we have studied the properties of light variations and the physical nature of the system. The spectroscopy reveals a spectral type of A3 for V1224 Cas. The light curves in both filters closely resemble those of the EL CVn-type systems with multi-periodic pulsations. We analyzed the eclipsing light curves by applying the W-D method. The photometric solution suggests that V1224 Cas is in detached state with a very small mass ratio of $\sim$0.087. The derived physical parameters in Table 2 indicate that the primary component of V1224 Cas is somewhat evolved, over-luminous, and over-sized, but still resides within the main-sequence band after a comparison of the mass-radius and mass-luminosity diagrams\citep{lba2006}. On the contrary, the secondary star is highly evolved, remarkably oversized, and over-luminous. Based on a mass of $\sim$0.19$M_{\sun}$, the secondary component of V1224 Cas can be thought to be a low-mass pre-He-WD. V1224 Cas can be further classified as a new EL CVn-type binary candidate consisting of a low-mass pre-He-WD and \textbf{an} A-type star. 

The light curves of V1224 Cas shows multi-periodic pulsations in addition to the eclipse-like light changes. The photometric solutions reveal that the less-massive secondary component is quite faint and contributes only $\sim$7\% to the total luminosity of the system. Accordingly, the short-term light variations can be put down to the intrinsic pulsations of the A-type primary star rather than the low-mass pre-WD component, since it cannot cause such pulsations with a semi-amplitude as high as $\sim$30 mmag. Besides, it can be seen from Figure 2 that the pulsations at the deeper primary light minimum appear to be significant than of that at the slightly shallower secondary light minimum. This indicates that the deeper eclipse could be attributed to the eclipse of the hotter pre-WD star by the A-type primary component, which in turn supports an EL CVn-type system for V1224 Cas. 

After subtracting the binary effects from the original observational data, we carried out a Fourier analysis of the light residuals to investigate the pulsation properties in detail. It leads to the detection of five confident pulsating frequencies in both $B$- and $V$-band data, including two eigen modes ($f_{1}$=15.4516 c/d, $f_{2}$=19.2351 c/d) and three probable harmonics. The ratios of the pulsational to the orbital periods are determined to be 0.028 and 0.023, both of which are within the upper limit of 0.09 for $\delta$ Sct stars in binaries \citep{zha2013}. Following \cite{zha2013}, the mean  density of the pulsating primary component is calculated to be $\rho_{1}$/$\rho_{\sun}$=0.049. Based on the well-known equation $Q=P_{pul}(\rho_{1}/\rho_{\sun})^{1/2}$, the pulsation constants of $f_{1}$ and $f_{2}$ are computed to be 0.0143 and 0.0115 days, respectively, corresponding to $p$ modes of $\delta$ Sct stars. The behaviors of the light curves, the period ratios, and the pulsation constants all suggest that the primary component of V1224 Cas is a $\delta$ Sct variable. As a conclusion, V1224 Cas could be a new EL CVn-type binary consisting a low-mass pre-He-WD and a $\delta$ Sct pulsator. 
 
\acknowledgments 
The authors are grateful to the anonymous referee for valuable comments. K.W acknowledges funding by the China Scholarship Council, the Meritocracy Research Funds of China West Normal University, and the Fundamental Research Funds of China West Normal University. C.Q.L is supported by Beijing Natural Science Foundation (grant 1184018). This work is supported by the National Natural Science Foundation of China (grants 11833002, 11633005, 11473037, and 11373037). The authors thank the Delingha site of Purple Mountain Observatory for its continuous support since 2009. They are grateful to all members of the site team, especially to the night assistants.

\textbf{\software{Period04 v.1.2\footnote{https://www.univie.ac.at/tops/Period04/} \citep{len2005}, IRAF \citep{tod1986,tod1993}, Wilson-Devinney (W-D) binary code\footnote{ftp://ftp.astro.ufl.edu/pub/wilson/lcdc2013/} \citep{wil1971,wil1979,wil1990,wil2012b}, Matplotlib \citep{hun2007}}}

\bibliographystyle{apj}
\bibliography{apj-jour}

\begin{thebibliography}{54}
\expandafter\ifx\csname natexlab\endcsname\relax\def\natexlab#1{#1}\fi

\bibitem[{{Aerts} {et~al.}(2010){Aerts}, {Christensen-Dalsgaard}, \&
  {Kurtz}}]{aer2010}
{Aerts}, C., {Christensen-Dalsgaard}, J., \& {Kurtz}, D.~W. 2010,
  {Asteroseismology}

\bibitem[{{Althaus} {et~al.}(2013){Althaus}, {Miller Bertolami}, \&
  {C{\'o}rsico}}]{alt2013}
{Althaus}, L.~G., {Miller Bertolami}, M.~M., \& {C{\'o}rsico}, A.~H. 2013,
  \aap, 557, A19

\bibitem[{{Breger}(2000)}]{bre2000}
{Breger}, M. 2000, in Astronomical Society of the Pacific Conference Series,
  Vol. 210, Delta Scuti and Related Stars, ed. M.~{Breger} \& M.~{Montgomery},
  3

\bibitem[{{Breger} {et~al.}(1993){Breger}, {Stich}, {Garrido}, {Martin},
  {Jiang}, {Li}, {Hube}, {Ostermann}, {Paparo}, \& {Scheck}}]{bre1993}
{Breger}, M., {et~al.} 1993, \aap, 271, 482

\bibitem[{{Bressan} {et~al.}(2012){Bressan}, {Marigo}, {Girardi}, {Salasnich},
  {Dal Cero}, {Rubele}, \& {Nanni}}]{bre2012}
{Bressan}, A., {Marigo}, P., {Girardi}, L., {Salasnich}, B., {Dal Cero}, C.,
  {Rubele}, S., \& {Nanni}, A. 2012, \mnras, 427, 127

\bibitem[{{Carter} {et~al.}(2011){Carter}, {Rappaport}, \&
  {Fabrycky}}]{car2011}
{Carter}, J.~A., {Rappaport}, S., \& {Fabrycky}, D. 2011, \apj, 728, 139

\bibitem[{{Chen} {et~al.}(2017){Chen}, {Maxted}, {Li}, \& {Han}}]{che2017}
{Chen}, X., {Maxted}, P.~F.~L., {Li}, J., \& {Han}, Z. 2017, \mnras, 467, 1874

\bibitem[{{Deng} {et~al.}(2013){Deng}, {Xin}, {Zhang}, {Li}, {Jiang}, {Wang},
  {Wang}, {Zhou}, {Yan}, \& {Luo}}]{den2013}
{Deng}, L., {et~al.} 2013, in IAU Symposium, Vol. 288, IAU Symposium, ed. M.~G.
  {Burton}, X.~{Cui}, \& N.~F.~H. {Tothill}, 318--319

\bibitem[{{Faigler} {et~al.}(2015){Faigler}, {Kull}, {Mazeh}, {Kiefer},
  {Latham}, \& {Bloemen}}]{fai2015}
{Faigler}, S., {Kull}, I., {Mazeh}, T., {Kiefer}, F., {Latham}, D.~W., \&
  {Bloemen}, S. 2015, \apj, 815, 26

\bibitem[{{Fan} {et~al.}(2016){Fan}, {Wang}, {Jiang}, {Wu}, {Li}, {Huang},
  {Xu}, {Hu}, {Zhu}, {Wang}, {Komossa}, \& {Zhang}}]{fan2016}
{Fan}, Z., {et~al.} 2016, \pasp, 128, 115005

\bibitem[{{Flower}(1996)}]{flo1996}
{Flower}, P.~J. 1996, \apj, 469, 355

\bibitem[{{Gaia Collaboration} {et~al.}(2018){Gaia Collaboration}, {Brown},
  {Vallenari}, {Prusti}, {de Bruijne}, {Babusiaux}, {Bailer-Jones}, {Biermann},
  {Evans}, {Eyer}, \& et~al.}]{bro2018}
{Gaia Collaboration} {et~al.} 2018, \aap, 616, A1

\bibitem[{{Gilliland} \& {Brown}(1988)}]{gil1988}
{Gilliland}, R.~L., \& {Brown}, T.~M. 1988, \pasp, 100, 754

\bibitem[{{Guo} {et~al.}(2017){Guo}, {Gies}, {Matson}, {Garc{\'{\i}}a
  Hern{\'a}ndez}, {Han}, \& {Chen}}]{guo2017}
{Guo}, Z., {Gies}, D.~R., {Matson}, R.~A., {Garc{\'{\i}}a Hern{\'a}ndez}, A.,
  {Han}, Z., \& {Chen}, X. 2017, \apj, 837, 114

\bibitem[{Hunter(2007)}]{hun2007}
Hunter, J.~D. 2007, Computing in Science Engineering, 9, 90

\bibitem[{{Ibano{\v g}lu} {et~al.}(2006){Ibano{\v g}lu}, {Soydugan},
  {Soydugan}, \& {Dervi{\c s}o{\v g}lu}}]{lba2006}
{Ibano{\v g}lu}, C., {Soydugan}, F., {Soydugan}, E., \& {Dervi{\c s}o{\v g}lu},
  A. 2006, \mnras, 373, 435

\bibitem[{{Istrate} {et~al.}(2016){Istrate}, {Marchant}, {Tauris}, {Langer},
  {Stancliffe}, \& {Grassitelli}}]{ist2016}
{Istrate}, A.~G., {Marchant}, P., {Tauris}, T.~M., {Langer}, N., {Stancliffe},
  R.~J., \& {Grassitelli}, L. 2016, \aap, 595, A35

\bibitem[{{Lee} {et~al.}(2017){Lee}, {Hong}, {Kim}, \& {Koo}}]{lee2017}
{Lee}, J.~W., {Hong}, K., {Kim}, S.-L., \& {Koo}, J.-R. 2017, \apj, 835, 189

\bibitem[{{Lenz} \& {Breger}(2005)}]{len2005}
{Lenz}, P., \& {Breger}, M. 2005, Communications in Asteroseismology, 146, 53

\bibitem[{{Liakos} \& {Niarchos}(2017)}]{lia2017}
{Liakos}, A., \& {Niarchos}, P. 2017, \mnras, 465, 1181

\bibitem[{{Lucy}(1967)}]{luc1967}
{Lucy}, L.~B. 1967, \zap, 65, 89

\bibitem[{{Luo} {et~al.}(2017){Luo}, {Wang}, {Zhang}, {Deng}, {Luo}, \&
  {Luo}}]{luo2017}
{Luo}, X., {Wang}, K., {Zhang}, X., {Deng}, L., {Luo}, Y., \& {Luo}, C. 2017,
  \aj, 154, 99

\bibitem[{{Marsh} {et~al.}(1995){Marsh}, {Dhillon}, \& {Duck}}]{mar1995}
{Marsh}, T.~R., {Dhillon}, V.~S., \& {Duck}, S.~R. 1995, \mnras, 275, 828

\bibitem[{{Maxted} {et~al.}(2014{\natexlab{a}}){Maxted}, {Serenelli}, {Marsh},
  {Catal{\'a}n}, {Mahtani}, \& {Dhillon}}]{max2014b}
{Maxted}, P.~F.~L., {Serenelli}, A.~M., {Marsh}, T.~R., {Catal{\'a}n}, S.,
  {Mahtani}, D.~P., \& {Dhillon}, V.~S. 2014{\natexlab{a}}, \mnras, 444, 208

\bibitem[{{Maxted} {et~al.}(2013){Maxted}, {Serenelli}, {Miglio}, {Marsh},
  {Heber}, {Dhillon}, {Littlefair}, {Copperwheat}, {Smalley}, {Breedt}, \&
  {Schaffenroth}}]{max2013}
{Maxted}, P.~F.~L., {et~al.} 2013, \nat, 498, 463

\bibitem[{{Maxted} {et~al.}(2014{\natexlab{b}}){Maxted}, {Bloemen}, {Heber},
  {Geier}, {Wheatley}, {Marsh}, {Breedt}, {Sebastian}, {Faillace}, {Owen},
  {Pulley}, {Smith}, {Kolb}, {Haswell}, {Southworth}, {Anderson}, {Smalley},
  {Collier Cameron}, {Hebb}, {Simpson}, {West}, {Bochinski}, {Busuttil}, \&
  {Hadigal}}]{max2014a}
---. 2014{\natexlab{b}}, \mnras, 437, 1681

\bibitem[{{Mkrtichian} {et~al.}(2004){Mkrtichian}, {Kusakin}, {Rodriguez},
  {Gamarova}, {Kim}, {Kim}, {Lee}, {Youn}, {Kang}, {Olson}, \&
  {Grankin}}]{mkr2004}
{Mkrtichian}, D.~E., {et~al.} 2004, \aap, 419, 1015

\bibitem[{{Mochejska} \& {Kaluzny}(1999)}]{noc1999}
{Mochejska}, B.~J., \& {Kaluzny}, J. 1999, \actaa, 49, 351

\bibitem[{{Montgomery} \& {Odonoghue}(1999)}]{mon1999}
{Montgomery}, M.~H., \& {Odonoghue}, D. 1999, Delta Scuti Star Newsletter, 13,
  28

\bibitem[{{Pollacco} {et~al.}(2006){Pollacco}, {Skillen}, {Collier Cameron},
  {Christian}, {Hellier}, {Irwin}, {Lister}, {Street}, {West}, {Anderson},
  {Clarkson}, {Deeg}, {Enoch}, {Evans}, {Fitzsimmons}, {Haswell}, {Hodgkin},
  {Horne}, {Kane}, {Keenan}, {Maxted}, {Norton}, {Osborne}, {Parley}, {Ryans},
  {Smalley}, {Wheatley}, \& {Wilson}}]{pol2006}
{Pollacco}, D.~L., {et~al.} 2006, \pasp, 118, 1407

\bibitem[{{Rappaport} {et~al.}(2015){Rappaport}, {Nelson}, {Levine},
  {Sanchis-Ojeda}, {Gandolfi}, {Nowak}, {Palle}, \& {Prsa}}]{rap2015}
{Rappaport}, S., {Nelson}, L., {Levine}, A., {Sanchis-Ojeda}, R., {Gandolfi},
  D., {Nowak}, G., {Palle}, E., \& {Prsa}, A. 2015, \apj, 803, 82

\bibitem[{{Rowe} {et~al.}(2010){Rowe}, {Borucki}, {Koch}, {Howell}, {Basri},
  {Batalha}, {Brown}, {Caldwell}, {Cochran}, {Dunham}, {Dupree}, {Fortney},
  {Gautier}, {Gilliland}, {Jenkins}, {Latham}, {Lissauer}, {Marcy}, {Monet},
  {Sasselov}, \& {Welsh}}]{row2010}
{Rowe}, J.~F., {et~al.} 2010, \apjl, 713, L150

\bibitem[{{Ruci{\'n}ski}(1969)}]{ruc1969}
{Ruci{\'n}ski}, S.~M. 1969, \actaa, 19, 245

\bibitem[{{S{\'a}nchez-Bl{\'a}zquez} {et~al.}(2006){S{\'a}nchez-Bl{\'a}zquez},
  {Peletier}, {Jim{\'e}nez-Vicente}, {Cardiel}, {Cenarro},
  {Falc{\'o}n-Barroso}, {Gorgas}, {Selam}, \& {Vazdekis}}]{san2006}
{S{\'a}nchez-Bl{\'a}zquez}, P., {et~al.} 2006, \mnras, 371, 703

\bibitem[{{Stellingwerf}(1978)}]{ste1978}
{Stellingwerf}, R.~F. 1978, \apj, 224, 953

\bibitem[{{Stetson}(2000)}]{ste2000}
{Stetson}, P.~B. 2000, \pasp, 112, 925

\bibitem[{{Tian} {et~al.}(2016){Tian}, {Deng}, {Zhang}, {Lu}, {Sun}, {Liu},
  {Zhou}, {Yan}, {Xin}, {Wang}, {Jiang}, {Luo}, \& {Yang}}]{tia2016}
{Tian}, J.~F., {et~al.} 2016, \pasp, 128, 105003

\bibitem[{{Tody}(1986)}]{tod1986}
{Tody}, D. 1986, in \procspie, Vol. 627, Instrumentation in astronomy VI, ed.
  D.~L. {Crawford}, 733

\bibitem[{{Tody}(1993)}]{tod1993}
{Tody}, D. 1993, in Astronomical Society of the Pacific Conference Series,
  Vol.~52, Astronomical Data Analysis Software and Systems II, ed. R.~J.
  {Hanisch}, R.~J.~V. {Brissenden}, \& J.~{Barnes}, 173

\bibitem[{{Torres} {et~al.}(2010){Torres}, {Andersen}, \&
  {Gim{\'e}nez}}]{tor2010}
{Torres}, G., {Andersen}, J., \& {Gim{\'e}nez}, A. 2010, \aapr, 18, 67

\bibitem[{{van Hamme}(1993)}]{van1993}
{van Hamme}, W. 1993, \aj, 106, 2096

\bibitem[{{van Kerkwijk} {et~al.}(2010){van Kerkwijk}, {Rappaport}, {Breton},
  {Justham}, {Podsiadlowski}, \& {Han}}]{van2010}
{van Kerkwijk}, M.~H., {Rappaport}, S.~A., {Breton}, R.~P., {Justham}, S.,
  {Podsiadlowski}, P., \& {Han}, Z. 2010, \apj, 715, 51

\bibitem[{{van Roestel} {et~al.}(2018){van Roestel}, {Kupfer}, {Ruiz-Carmona},
  {Groot}, {Prince}, {Burdge}, {Laher}, {Shupe}, \& {Bellm}}]{van2018}
{van Roestel}, J., {et~al.} 2018, \mnras, 475, 2560

\bibitem[{{Wang} {et~al.}(2015{\natexlab{a}}){Wang}, {Zhang}, {Deng}, {Luo},
  {Luo}, \& {Zhang}}]{wan2015a}
{Wang}, K., {Zhang}, X., {Deng}, L., {Luo}, C., {Luo}, Y., \& {Zhang}, J.
  2015{\natexlab{a}}, \apj, 805, 22

\bibitem[{{Wang} {et~al.}(2015{\natexlab{b}}){Wang}, {Deng}, {Zhang}, {Xin},
  {Yan}, {Tian}, {Luo}, {Luo}, {Zhang}, {Peng}, {Pan}, {Sun}, \&
  {Luo}}]{wan2015b}
{Wang}, K., {et~al.} 2015{\natexlab{b}}, \aj, 150, 161

\bibitem[{{Wilson}(1979)}]{wil1979}
{Wilson}, R.~E. 1979, \apj, 234, 1054

\bibitem[{{Wilson}(1990)}]{wil1990}
---. 1990, \apj, 356, 613

\bibitem[{{Wilson}(2012)}]{wil2012b}
---. 2012, Journal of Astronomy and Space Sciences, 29, 115

\bibitem[{{Wilson} \& {Devinney}(1971)}]{wil1971}
{Wilson}, R.~E., \& {Devinney}, E.~J. 1971, \apj, 166, 605

\bibitem[{{Wu} {et~al.}(2007){Wu}, {Zhou}, {Ma}, {Jiang}, {Chen}, \&
  {Wu}}]{wu2007}
{Wu}, Z.-Y., {Zhou}, X., {Ma}, J., {Jiang}, Z.-J., {Chen}, J.-S., \& {Wu},
  J.-H. 2007, \aj, 133, 2061

\bibitem[{{Zhang} {et~al.}(2016){Zhang}, {Fu}, {Li}, {Ren}, \& {Luo}}]{zha2016}
{Zhang}, X.~B., {Fu}, J.~N., {Li}, Y., {Ren}, A.~B., \& {Luo}, C.~Q. 2016,
  \apjl, 821, L32

\bibitem[{{Zhang} {et~al.}(2017){Zhang}, {Fu}, {Liu}, {Luo}, \&
  {Ren}}]{zha2017}
{Zhang}, X.~B., {Fu}, J.~N., {Liu}, N., {Luo}, C.~Q., \& {Ren}, A.~B. 2017,
  \apj, 850, 125

\bibitem[{{Zhang} {et~al.}(2013){Zhang}, {Luo}, \& {Fu}}]{zha2013}
{Zhang}, X.~B., {Luo}, C.~Q., \& {Fu}, J.~N. 2013, \apj, 777, 77

\bibitem[{{Zhang} {et~al.}(2015){Zhang}, {Luo}, \& {Wang}}]{zha2015}
{Zhang}, X.~B., {Luo}, Y.~P., \& {Wang}, K. 2015, \aj, 149, 96

\end{thebibliography}

\end{document}